\documentclass[superscriptaddress,amsmath,amssymb,nofootinbib,eqsecnum,a4paper,secnumarabic,11pt]{revtex4}
\usepackage{ascmac,braket,bm,mathrsfs,amsthm,amsfonts}
\usepackage{hyperref}
\usepackage{graphicx}
\usepackage{multirow}
\usepackage{titlesec}
\usepackage{ulem}
\newcommand*{\justifyheading}{\raggedright}
\titleformat{\chapter}[display]
  {\normalfont\huge\bfseries\justifyheading}{\chaptertitlename\ \thechapter}
  {20pt}{\Huge}
\titleformat{\section}
  {\normalfont\Large\bfseries\justifyheading}{\thesection}{1em}{}
\titleformat{\subsection}
  {\normalfont\large\bfseries\justifyheading}{\thesubsection}{1em}{}
\titleformat{\subsubsection}
  {\normalfont\bfseries\justifyheading}{\thesubsubsection}{1em}{}
\usepackage[english]{babel}
\usepackage[dvipsnames]{xcolor}
\newcommand{\gsim}{\raisebox{-0.13cm}{~\shortstack{$>$ \\[-0.07cm]
      $\sim$}}~}
\newcommand{\lsim}{\raisebox{-0.13cm}{~\shortstack{$<$ \\[-0.07cm]
      $\sim$}}~}

\begin{document}

\begin{flushright}
{\tt IPMU25-0007}
\end{flushright}

\title{Extension of SUSY $SU(5)$ GUTs with Nelson-Barr models}

\author{Junji Hisano}
\email{hisano@eken.phys.nagoya-u.ac.jp}
\affiliation{
  Kobayashi-Maskawa Institute for the Origin of Particles and the
  Universe, Nagoya University,
  Furo-cho Chikusa-ku, Nagoya, 464-8602 Japan
}
\affiliation{
  Department of Physics, Nagoya University, Furo-cho Chikusa-ku, Nagoya, 464-8602 Japan
}
\affiliation{
  Kavli IPMU (WPI), UTIAS, University of Tokyo, Kashiwa, 277-8584, Japan
}

\author{Moe Kuroda}
\affiliation{
  Department of Physics, Nagoya University, Furo-cho Chikusa-ku, Nagoya, 464-8602 Japan
}

\begin{abstract}
  In this paper we extend the supersymmetric $SU(5)$ GUTs with  the Nelson-Barr (NB) models. The NB models are a candidate for the solution of the strong CP problem. We show that the parameters in the CKM matrix are reproduced while the mass ratios of the down-type quarks and leptons in the second and third generations are explained in the minimal extension. 
\end{abstract}

\maketitle

\clearpage
\tableofcontents

\section{Introduction}

The strong $CP$ problem is one of the important issues when constructing models beyond the Standard Model (BSM). The QCD $\theta$ parameter is constrained by the null result in the measurement of the neutron electric dipole moment as $\theta \lesssim 10^{-10}$, while the $CP$ phase in the Cabibbo-Kobayashi-Maskawa (CKM) matrix is $\mathcal{O}(1)$ in the Standard Model (SM). This naturalness problem is considered difficult to solve using the Anthropic Principle \cite{Weinberg:1987dv,Bousso:2000xa,Susskind:2003kw,Kachru:2003aw}. Proposals for this problem are sometimes combined with other particle physics models.

The Nelson-Barr (NB) models \cite{Nelson:1983zb,Barr:1984qx, Barr:1984fh} are a candidate for solving the strong $CP$ problem. In the models, $CP$ symmetry is assumed to be spontaneously broken at a high-energy scale. The models are designed so that the determinant of the colored fermion mass matrix does not have a phase even after $CP$ symmetry is broken. To introduce the $CP$ phase in the CKM matrix, vector-like quarks are introduced and mixed with SM quarks using complex coefficients.

In this paper, we introduce the NB models into the supersymmetric $SU(5)$ grand unified models (SUSY $SU(5)$ GUTs)  \cite{Witten:1981nf,Dimopoulos:1981yj,Dimopoulos:1981zb,Sakai:1981gr}. SUSY $SU(5)$ GUTs are a strong candidate for BSM due to the  gauge coupling unification \cite{Langacker:1991an,Amaldi:1991cn}. However, the mass ratios of down-type quarks and charged leptons in each generation, which result from common Yukawa couplings, are predicted by the models but are not consistent with observed values. We demonstrate that the minimal extension of SUSY $SU(5)$ GUTs with the NB models, which introduces a vector-like chiral multiplet, makes the mass ratios of the second and third generations consistent with observations while also accounting for the CKM matrix.

SUSY has pros and cons for the NB models. The non-renormalization theorem suppresses higher-order contributions to the QCD $\theta$ parameter \cite{Ellis:1982tk}. Since the constraint on this parameter is stringent, even higher-order contributions must be suppressed in non-SUSY cases\footnote{
The higher-order contributions to the QCD $\theta$ parameters are discussed in Ref.~\cite{Hisano:2023izx,Banno:2023yrd,Banno:2025pfq}.}  \cite{Dine:2015jga}. On the other hand, SUSY is not exact in nature, necessitating the introduction of SUSY breaking terms in the SUSY SM, which might contribute to the QCD $\theta$ parameter if they are complex \cite{Dine:1993qm}. Some SUSY-breaking mediation models, such as gauge \cite{Dine:1994vc} and anomaly mediation \cite{Randall:1998uk,Giudice:1998xp}, could help address this issue. The recent discuss is found in Ref.~\cite{Evans:2020vil}.

This paper is organized as follows. In Sec.~2, we review the NB models and discuss the favorable parameter space for explaining the CKM phase. In Sec.~3, we introduce the NB models to SUSY $SU(5)$ GUTs and show that the mass ratios of down-type quarks and charged leptons in the second and third generations are economically explained, in addition to the CKM matrix, within this extended framework. Sec.~4 is devoted to concludes.

\section{Nelson-Barr Models}

In the SM, the Yukawa couplings for quarks are given as 
\begin{eqnarray}
  - {\cal L}_{\rm SM}&=&
  \bar{y}^u_{ij} (\bar{Q}_i\tilde{H}) u_j+  \bar{y}^d_{ij}(\bar{Q}_i H) d_i +{\rm h.c.}
\end{eqnarray}
where $i,j=1$-3. Here, $H$ is the doublet Higgs boson ($\tilde{H}\equiv \epsilon H^\star$), and $Q_i$,$u_i$, and $d_i$ are for doublet and singlet quarks, respectively. In a basis where $\bar{y}^d_{ij}$ is diagonal ($\bar{y}^d_{ij}=\bar{y}^d_i\delta_{ij}$),  $\bar{y}^u_{ij}=\bar{V}^\dagger_{ij}\bar{y}_j^u$ ($\bar{V}$ is the CKM matrix in the SM). The vector-like quarks are introduced in the NB models
in order to generate the $CP$ phase in the CKM matrix without generating the QCD $\theta$ parameter. In this paper, we introduce a vector-like quark $D$ as the simplest model,  whose gauge quantum numbers are the same as the singlet down-type quarks ($d$-mediation) \cite{Bento:1991ez}. We could also introduce a vector-like quark $U$ with the gauge quantum numbers of  the singlet up-type quarks ($u$-mediation).  In this section, we will review the $d$-mediation NB model since we extend the SUSY $SU(5)$ GUTs using the model in next section.

The vector-like quark has couplings\footnote{
The complex conjugates of the scalar fields, $\Sigma_a^\star$, are here, since the chiral superfields include  $\Sigma_a$ in next section.
} with complex scalar fields $\Sigma_a$ ($a=1\cdots N$) as $f_{ai} \bar{D}_L d_j \Sigma_a^\star$.  We assume that the $CP$ symmetry is spontaneously broken by $\langle \Sigma_a\rangle \ne 0$. $N$ is larger than 1 so that the phases cannot be removed by the rephasing of fields.  Parameterizing the vacuum expectation values (VEVs) of $\Sigma_a$ as $f_{ai}\langle \Sigma_a\rangle = \xi_i {\rm e}^{i\varphi_i}$, we get
\begin{eqnarray}
  - {\cal L}_{\rm NB}&=&
  V^{uT}_{ij}y_j^u(\bar{Q}_i\tilde{H}) u_j+  y^d_i(\bar{Q}_i H) d_i
+\xi_j{\rm e}^{-i \varphi_j}\bar{D}_Ld_j+m \bar{D}_LD_R
  +{\rm h.c.}.
\end{eqnarray}
It is assumed that  discrete symmetries forbid the terms contributing to the QCD $\theta$ parameter, such as $\Sigma_a \bar{D}_L D_R$. Here, $V^u$ is a real orthogonal matrix. The Yukawa coupling constants and mass, $y_i^u$, $y_i^u$,  and $m$ are real. After rephasing the quark fields as $Q_i\rightarrow {\rm e}^{i\varphi_i} Q_i$, $u_i\rightarrow {\rm e}^{i\varphi_i} u_i$, and $d_i\rightarrow {\rm e}^{i\varphi_i} d_i$, we get 
\begin{eqnarray}
  - {\cal L}_{\rm NB}&=&
  {\rm e}^{-i(\varphi_i-\varphi_j)} V^{uT}_{ij}y_j^u(\bar{Q}_i\tilde{H}) u_j+  y^d_i(\bar{Q}_i H) d_i
+\xi_j\bar{D}_Ld_j+m \bar{D}_LD_R
  +{\rm h.c.}
\end{eqnarray}
This rephasing leaves the QCD $\theta$ parameter invariant.
 Assuming that the vector-like quark mass $M$ ($=\sqrt{m^2+\xi^2}$ with $\xi^2=\sum_i\xi_i^2$) is heavy enough,
 the SM down-type quark Yukawa couplings are given after integrating out it as \cite{Valenti:2021rdu}
\begin{eqnarray}
  \bar{y}^d_{ij} &=& y^d_i\left(
  \delta_{ij}-\frac{\xi_i \xi_j}{\xi^2}
  (1-\frac{m}{M})
  \right).
\end{eqnarray}

When the Yukawa couplings are parameterized as $\bar{y}^d_{ij}=V^{dT}_{ik}\bar{y}^d_k U^{d}_{kj}$, the CKM matrix is given as
\begin{eqnarray}
  \bar{V}_{ij}&=&  V^u_{ik} {\rm e}^{-i(\varphi_i-\varphi_k)} V^{dT}_{kj}.
\label{eq:ckm_in_NB}
\end{eqnarray}

Now we assume $\xi_1=0$ (then, $\xi_2=\xi \cos\theta$ and $\xi_3=\xi\sin\theta$) in order to  evaluate the SM quark Yukawa couplings  in a semi-analytic way. In this case, $V^d$ is given as
\begin{eqnarray}
  V^d&=&\left(\begin{array}{ccc}
    1&&\\&\cos\psi&\sin\psi\\&-\sin\psi&\cos\psi
  \end{array}\right)
\end{eqnarray}
Taking $\bar{y}^d_3\gg\bar{y}^d_2$, we obtain the approximate formulae
\begin{eqnarray}
  (\bar{y}_3^d)^2&\simeq& (y^d_3)^2(1-(2-A)A \sin^2\theta),\nonumber\\
  (\bar{y}_2^d)^2&\simeq& (y^d_2)^2 \frac{(1-A)^2}{(1-(2-A)A \sin^2\theta)},\nonumber\\
 \tan2 \psi&\simeq& \frac{(2-A)A\sin2\theta}{1-A}\frac{\bar{y}_2^d}{\bar{y}_3^d}.
\end{eqnarray}
where $A=(1-m/M)$ ($0<A<1$). When $\xi\gsim m$ $(A\simeq 1-m/\xi)$,
\begin{eqnarray}
 (\bar{y}_3^d)^2&\simeq& (y^d_3)^2\cos^2\theta,\nonumber\\
  (\bar{y}_2^d)^2&\simeq& (y^d_2)^2 \frac{(m/\xi)^2}{\cos^2\theta},\nonumber\\
 \tan2 \psi&\simeq& \frac{\bar{y}_2^d}{\bar{y}_3^d} (\xi/m) \sin2\theta.
\end{eqnarray}  
Then, the bottom and strange Yukawa coupling constants become smaller than without the mixing with the vector-like quark, assuming $(m/\xi)^2\lsim \cos^2\theta$.  On the other hand, $\xi\lsim m$ ($A\simeq \xi^2/(2m^2)$), 
\begin{eqnarray}
 (\bar{y}_3^d)^2&\simeq& (y^d_3)^2,\nonumber\\
  (\bar{y}_2^d)^2&\simeq& (y^d_2)^2,\nonumber\\
  \tan2 \psi&\simeq&  \frac{\bar{y}_2^d}{\bar{y}_3^d}(\xi/m)^2 \sin 2\theta.
\end{eqnarray}  
The bottom and strange Yukawa coupling constants are not suppressed by the mixing with the vector-like quark\footnote{
These results are  reasonable since  ${\rm det}(\bar{y}^d)={\rm det}({y}^d)\times (m/M)$.}.

Next, we consider the CKM matrix in the NB model.  The Jarslskog invariant $J$ of the CKM matrix  $\bar{V}$ is
\begin{eqnarray}
  {\rm Im}[\bar{V}_{ij}\bar{V}_{kl}\bar{V}_{il}^\star \bar{V}_{kj}^\star]=J\sum_{mn}
    \epsilon_{ikm}\epsilon_{jln},
\end{eqnarray}
and $J\sim O(\lambda^6)$ in the Wolfenstein parameterization with $\lambda\simeq 0.22$. When Eq.~(\ref{eq:ckm_in_NB}) is inserted into $J$, we get
\begin{eqnarray}
  J&=& V^u_{11}V^u_{21}(-V^u_{13}V^u_{22}+V^u_{12}V^u_{23})\sin\psi\cos\psi\sin(\varphi_3-\varphi_2),
\end{eqnarray}  
When the magnitudes of the components in $V^u$ are comparable to those in the observed CKM matrix, such as $V^u_{12}\sim \lambda$, $V^u_{13}\sim \lambda^3$, and $V^u_{23}\sim \lambda^2$, it would be easier to reproduce the CKM matrix in the NB model. In the case, 
$\sin\psi \sim \lambda^2$ is required from $J\sim O(\lambda^6)$  under an assumption $\sin(\varphi_3-\varphi_2)=O(1)$. This is consistent with the above assumption that $V^u$ has a similar structure to the observed CKM matrix. Using Table~\ref{tab:mytable1}, we get $\bar{y}^d_3/\bar{y}^d_2\sim 55$ at the $m_Z$ scale. the ratio is not so changed even at the higher energy scale.  Using the value, it is found that $\xi \gsim m$ and  $(\xi/m)\sin\theta\simeq(2\mbox{-}3)$. This implies that the bottom and strange quark masses are suppressed by the mixing with the vector-like quark.

\begin{table}[h] 
\begin{tabular}{|c|c|c|c|c|c|c|}
\hline
\multirow{3}{*}{fermion masses}
& $m_t(m_Z)$  & 168.26 GeV & $m_b(m_Z)$  & 2.839 GeV & $m_{\tau}(m_Z)$  & 1.72856 GeV \\ \cline{2-7}
& $m_c(m_Z)$  & 0.620 GeV & $m_s(m_Z)$  & 53.16 MeV & $m_{\mu}(m_Z)$   & 0.101766 GeV \\ \cline{2-7}
& $m_u(m_Z)$  & 1.23 MeV & $m_d(m_Z)$  & 2.67 MeV & $m_e(m_Z)$  & 0.48307 MeV\\ \hline
gauge couplings & $\alpha_1(m_Z)$ & 0.0169 & $\alpha_2(m_Z)$ & 0.0338 & $\alpha_3(m_Z)$ & 0.1187 \\ \hline
\multirow{2}{*}{CKM matrix}& $\sin{\theta_{12}}$ & 0.22501 & $\sin{\theta_{23}}$ & 0.04183 \\ \cline{2-5}
&$\sin\theta_{13}$ & 0.003732 & $\delta $ & 0.147 \\ \cline{1-5}
Higgs 4-pt coupling &0.13947\\ \cline{1-2}
\end{tabular}
\caption{Parameters used in this paper. Fermion masses, gauge coupling constants, and Higgs four-pint coupling in the SM are $\overline{\rm MS}$ parameters at $Z$ boson mass ($m_Z$) scale, given in Ref.~\cite{Huang:2020hdv}. Parameters in CKM matrix comes from PDG \cite{ParticleDataGroup:2024cfk}. 
}
\label{tab:mytable1} %
\end{table}

Here, we assumed $\xi_1=0$, but a similar analysis applies if  $\xi_2=0$ or  $\xi_3=0$. In those cases, $\xi/m$ is also larger than one. This is consistent with Ref.~\cite{Valenti:2021rdu}. In any cases, the down-type quarks in the SM have smaller than without the mixing with the vector-like quark.

We will apply the above result to reproduce the mass ratios between bottom quark and tau lepton and between strange quark and muon in the SUSY $SU(5)$ GUTs by introducing the NB models.

%
\section{Extension of SUSY $SU(5)$ GUTs with NB models}
\label{sec:SUSYGUT}

Now we consider extension of the minimal SUSY $SU(5)$ GUT with the NB model. In the minimal SUSY $SU(5)$ GUTs, the doublet quarks ($Q_i$), the singlet up quarks ($u_i^c$) and the singlet leptons $(e_i^c)$ are embedded into {\bf 10}-dimensional multiplets ($\psi_i$),  and the singlet down quarks $(d_i^c)$ and the doublet leptons $(L_i)$ are into {\bf 5$^\star$}-dimensional multiplets ($\phi_i$). The Yukawa couplings of quarks and leptons are given by the following superpotential,
\begin{eqnarray}
  W_{SU(5)}&=&\frac{h^{ij}}{4}\epsilon_{\alpha\beta\gamma\delta\epsilon}
  \psi_i^{\alpha\beta}\psi_j^{\gamma\delta}H^\epsilon+\sqrt{2} f^{ij}\psi^{\alpha\beta}_i \phi_{j\alpha}\bar{H}_\beta
\end{eqnarray}
where $\alpha,\beta,\cdots=1\cdots 5$. $H$ and $\bar{H}$ are   {\bf 5} and {\bf 5$^\star$}-dimensional multiplets, which includes the doublet Higgses, $H_u$ and $H_d$, in the SUSY SM. The Yukawa couplings are decomposed as
\begin{eqnarray}
  h^{ij}&=&V^{uT}_{ik}y^u_k{\rm e}^{i\varphi_k} V^u_{kj}, \label{eq:orthogonal}\\
  f^{ij}&=&y^d_i\delta_{ij},
\end{eqnarray}
where $\varphi_i$ ($\sum_i\varphi_i=0$) are generic $CP$ phases in the SUSY $SU(5)$ GUTs \cite{Hisano:1992jj}.

The SUSY SM superpotential is derived from the model as 
\begin{eqnarray}
  W_{\rm SUSY~SM}&=&\bar{V}^T_{ij} \bar{y}^u_j Q_i u^c_j H_u+\bar{y}^d_i Q_i d^c_i H_d+\bar{y}^l_i e^c_i L_i H_d.
\end{eqnarray}
In the SUSY SU(5) GUTs, $\bar{V}=V^u$ and the Yukawa coupling constants for down-type quarks and leptons are the same for each generation at the GUT scale\footnote{
In this paper we take $M_G=2\times 10^{16}$~GeV.
}
($M_G$), $\bar{y}^d_i=\bar{y}^l_i(=y^d_i)$. On the other hand, their ratios at the GUT scale  evaluated from observables are
\begin{eqnarray}
  \frac{\bar{y}^d_1}{\bar{y}^l_1}\simeq 2.4,~~
  \frac{\bar{y}^d_2}{\bar{y}^l_2}\simeq 0.23,~~
\frac{\bar{y}^d_3}{\bar{y}^l_3}\simeq 0.76.
\end{eqnarray}
We evaluate them using the quark and lepton masses at the $m_Z$ scale in Table~\ref{tab:mytable1} with the renormalization-group equations at two-loop level \cite{Barger:1992ac}. Here, we assume that  the SUSY breaking scale in the SUSY SM ($m_{\rm SUSY}$) is 1~TeV. The ratio of ${\bar{y}^d_3}/{\bar{y}^l_3}$ is deviated from the above value when $\tan\beta$ is around 1 or larger than 60 due to the large top or bottom Yukawa couplings. The threshold correction to the Yukawa coupling constants at $m_{\rm SUSY}$ might bridge the gap between theoretical predictions and observations, since the corrections proportional to $\tan\beta$ may become $O(1)$.

In this paper, we try to explain the ratios of the second and third generations by introducing the NB model to the SUSY $SU(5)$ GUTs. We introduce a pair of {\bf 5} and {\bf 5$^\star$}-dimensional multiplets, $F$ and $\bar{F}$,  and $SU(5)$ singlets, $\Sigma_a$, whose vacuum expectation values are complex, and they have the following superpotential\footnote{More concrete models of extension of the SUSY $SU(5)$ GUTs with the NB model are given in Ref.~\cite{Evans:2020vil}.},
\begin{eqnarray}
  W^{(1)}_{SU(5)}&=& f_{ai} \Sigma_{a} F\phi_i + F(M+g \Phi)\bar{F}.
\end{eqnarray}
$F$ and $\bar{F}$ are coupled with the $SU(5)$ breaking Higgs $\Phi$ ({\bf 24}-dimensional multiplet) so that the $SU(3)_C$ triplets and  $SU(2)_L$ doublets in $F$ and $\bar{F}$ have different masses from each other ($m_3$ and $m_2$). When parameterizing the vacuum expectation values of $\Sigma_a$ as $f_{ai} \langle \Sigma_{a}\rangle = \xi_i {\rm e}^{i\varphi_i}$, the results in the previous section can be used here.

Now we would like to reproduce the mass ratios between bottom quark and tau lepton and also between strange quark and muon, but not one between down quark and electron. The ratios in the third and second generations are smaller than 1, while that of the fist generation is larger than 1. The ratio of the first generation might be explained by introduction of the higher-dimensional operators including $\Phi$, as the smaller Yukawa couplings are more susceptible to contributions from such operators. We assume $\xi_1=0$, $\xi_2= \xi\cos\theta$, and $\xi_3=\xi\sin\theta$ again. If $m_3 \lsim \xi$ and $m_2\gsim \xi$, we get
\begin{eqnarray}
  (\bar{y}_3^d)^2\simeq (\bar{y}_3^l)^2 \cos^2\theta,\\
  (\bar{y}_2^d)^2\simeq (\bar{y}_2^l)^2 \frac{(m_3/\xi)^2}{\cos^2\theta}.
\end{eqnarray}
Thus, we can explain the ratios of the second and third generations by choosing $\cos\theta$ and $m_3/\xi(\lsim \cos\theta)$.

Now we demonstrate the mass ratios of the second and third generations and also CKM matrix including the $CP$ phase can be explained in the extension with the NB model. We assume that $m_3$, $m_2$ and $\xi$ are the GUT scale for simplicity. The input parameters are $\xi/m_3$, $\theta$, $\varphi_3-\varphi_2$ and also three angles $\theta_i$ $(i=1-3)$ in the real orthogonal matrix $V^u$ in Eq.~(\ref{eq:orthogonal}),
\begin{eqnarray}
  V^u&=&
  \left(\begin{array}{ccc}
    \cos\theta_{3}&\sin\theta_{3}&\\
    -\sin\theta_{3}&\cos\theta_{3}&\\
    &&1
    \end{array}
    \right)
  \left(\begin{array}{ccc}
    \cos\theta_{2}&&-\sin\theta_{2}\\
    &1&\\
    \sin\theta_{2}&&\cos\theta_{2}\\
    \end{array}
    \right)
  \left(\begin{array}{ccc}
    1&&\\
    &\cos\theta_{1}&\sin\theta_{1}\\
    &-\sin\theta_{1}&\cos\theta_{1}\\
    \end{array}
    \right).
\end{eqnarray}
We assume that $m_2$ is heavy enough, and then it is decoupled from the parameter fitting.
Since we have six input parameters to fit six observables, two ratios and four parameters in the CKM matrix given in Table~\ref{tab:mytable1},  all parameters can be fixed if we can find the solutions. In fact, we found the solutions easily in a numerical way. We did not reply on the approximate formulae derived in the previous section. We found two solutions for $\tan\beta=3$, 10, 30 and $m_{\rm SUSY}=1$~TeV and 100~TeV with opposite sign of $\varphi_3-\varphi_2$. They are shown in  Table~\ref{tab:mytable2}.

\begin{table}
\centering 
\begin{tabular}{|c|c|c|}
\hline
$m_{\mathrm{SUSY}}$ & $\tan{\beta}$
&($|{\xi}/{m_{3}}|,\,\theta,\,\theta_1,\,\theta_2,\,\theta_3,\,\varphi_3-\varphi_2$) \\ \hline
\multirow{6}{*}{$1\,\mathrm{TeV}$} & 
\multirow{2}{*}{3} 
& $(5.62,\, 0.705,\, 0.0101,\, -0.00794,\, 0.227,\, 0.330)$ \\ \cline{3-3}
&& $(5.62,\, 0.705,\, 0.0797,\, 0.00794,\, 0.227,\, -0.330)$ \\ \cline{2-3}
& \multirow{2}{*}{10} 
& $(5.71,\, 0.722,\, 0.0111,\, -0.00805,\, 0.227,\, 0.324)$   \\ \cline{3-3}
&  & $(5.71,\, 0.722,\, 0.0817,\, 0.00805,\, 0.227,\,-0.324)$   \\ \cline{2-3}
& \multirow{2}{*}{30} 
& $(5.70, \,0.719,\, 0.00945,\, -0.00794,\, 0.227,\, 0.334)$   \\ \cline{3-3}
&  & $(5.70,\, 0.719,\, 0.0791,\, 0.00794,\, 0.227,\, -0.334)$   \\ \hline
\multirow{6}{*}{$100\,\mathrm{TeV}$} & 
\multirow{2}{*}{3}  
& $(5.80, \,0.757,\, 0.0130,\, -0.00844,\, 0.227,\, 0.315)$ \\ \cline{3-3}
&& $(5.80,\, 0.757, \,0.0870,\, 0.00844,\, 0.227,\, -0.315)$ \\ \cline{2-3}
& \multirow{2}{*}{10} 
& $(5.85, \,0.767,\, 0.0136,\, -0.00850,\, 0.227,\, 0.313)$   \\ \cline{3-3}
&  & $(5.85, \,0.767,\, 0.0881,\, 0.00850,\, 0.227,\, -0.313)$   \\ \cline{2-3}
& \multirow{2}{*}{30} 
& $(5.90, \,0.775, \,0.0130,\, -0.00842,\, 0.227,\, 0.316)$   \\ \cline{3-3}
&  & $(5.90, \,0.775,\, 0.0868,\, 0.00842,\, 0.227,\, -0.316)$   \\ \hline
\end{tabular}
\caption{Parameter sets to explain the mass ratios of the second and third generations and also four parameters in the CKM matrix for fixed $\tan{\beta}$ and $m_{\mathrm{SUSY}}$.}
\label{tab:mytable2} %
\end{table}

In this section we assumed the masses of the vector-like fermions are around the GUT scale. Then, we solved the renormalization group equations of the Yukawa couplings in the SUSY SM at two-loop level, while we ignored the threshold correction to the Yukawa couplings at $m_{\rm SUSY}$. We found the solution to explain the mass ratios of the second and third generations and also four parameters in the CKM matrix. We consider that even if the masses of the vector-like fermion can be much smaller than the GUT scale, we can explain those observables. 

Since $\xi/m_2 \lsim 1$ is assumed, the $CP$ phase in the Pontecorvo–Maki–Nakagawa–Sakata (PMNS) matrix might be smaller than $O(1)$ when we introduce the seesaw mechanism with the right-handed neutrinos. However, if the Yukawa couplings $\Sigma_a (\nu^c)^2$ ($\nu^c$ are for the right-handed neutrinos) are introduced in the superpotential, the $O(1)$ phase in the PMNS matrix might be possible \cite{Evans:2020vil}. 

\section{Conclusions}
\label{sec:conclusions}

In this paper we introduce the Nelson-Barr (NB) models to the SUSY $SU(5)$ GUTs. The NB models are  a candidate for the solution of the strong $CP$ problem. We show that the parameters in the CKM matrix are reproduced while the mass ratios of the down-type quarks and leptons in the second and third generations are explained in the extension. We assumed that the vector-like fermions have masses around the GUT scale for simplicity, though we can lower them from the GUT scale. This could be advantageous as it potentially alleviates the quality problem \cite{Asadi:2022vys}.

\section*{Acknowledgments}

\noindent
This work is supported by the JSPS Grant-in-Aid for Scientific Research Grant No.\,23K20232 (J.H.) and  No.\,24K07016 (J.H.). The work of J.H.\ is also supported by World Premier International Research Center Initiative (WPI Initiative), MEXT, Japan.
This work is also supported by JSPS Core-to-Core Program Grant No.\,JPJSCCA20200002. 


\appendix



{}


\begin{thebibliography}{99}

\bibitem{Weinberg:1987dv}
S.~Weinberg,
Phys. Rev. Lett. \textbf{59} (1987), 2607
doi:10.1103/PhysRevLett.59.2607

\bibitem{Bousso:2000xa}
R.~Bousso and J.~Polchinski,
JHEP \textbf{06} (2000), 006
doi:10.1088/1126-6708/2000/06/006
[arXiv:hep-th/0004134 [hep-th]].

\bibitem{Susskind:2003kw}
L.~Susskind,
[arXiv:hep-th/0302219 [hep-th]].

\bibitem{Kachru:2003aw}
S.~Kachru, R.~Kallosh, A.~D.~Linde and S.~P.~Trivedi,
Phys. Rev. D \textbf{68} (2003), 046005
doi:10.1103/PhysRevD.68.046005
[arXiv:hep-th/0301240 [hep-th]].


\bibitem{Nelson:1983zb}
A.~E.~Nelson,
Phys. Lett. B \textbf{136} (1984), 387-391
doi:10.1016/0370-2693(84)92025-2

\bibitem{Barr:1984qx}
S.~M.~Barr,
Phys. Rev. Lett. \textbf{53} (1984), 329
doi:10.1103/PhysRevLett.53.329

\bibitem{Barr:1984fh}
S.~M.~Barr,
Phys. Rev. D \textbf{30} (1984), 1805
doi:10.1103/PhysRevD.30.1805

\bibitem{Witten:1981nf}
E.~Witten,
Nucl. Phys. B \textbf{188} (1981), 513
doi:10.1016/0550-3213(81)90006-7

\bibitem{Dimopoulos:1981yj}
S.~Dimopoulos, S.~Raby and F.~Wilczek,
Phys. Rev. D \textbf{24} (1981), 1681-1683
doi:10.1103/PhysRevD.24.1681

\bibitem{Dimopoulos:1981zb}
S.~Dimopoulos and H.~Georgi,
Nucl. Phys. B \textbf{193} (1981), 150-162
doi:10.1016/0550-3213(81)90522-8

\bibitem{Sakai:1981gr}
N.~Sakai,
Z. Phys. C \textbf{11} (1981), 153
doi:10.1007/BF01573998


\bibitem{Langacker:1991an}
P.~Langacker and M.~x.~Luo,
Phys. Rev. D \textbf{44} (1991), 817-822
doi:10.1103/PhysRevD.44.817

\bibitem{Amaldi:1991cn}
U.~Amaldi, W.~de Boer and H.~Furstenau,
Phys. Lett. B \textbf{260} (1991), 447-455
doi:10.1016/0370-2693(91)91641-8

\bibitem{Dine:2015jga}
M.~Dine and P.~Draper,
JHEP \textbf{08} (2015), 132
doi:10.1007/JHEP08(2015)132
[arXiv:1506.05433 [hep-ph]].

\bibitem{Hisano:2023izx}
J.~Hisano, T.~Kitahara, N.~Osamura and A.~Yamada,
JHEP \textbf{03} (2023), 150
doi:10.1007/JHEP03(2023)150
[arXiv:2301.13405 [hep-ph]].

\bibitem{Banno:2023yrd}
T.~Banno, J.~Hisano, T.~Kitahara and N.~Osamura,
JHEP \textbf{02} (2024), 195
doi:10.1007/JHEP02(2024)195
[arXiv:2311.07817 [hep-ph]].

\bibitem{Banno:2025pfq}
T.~Banno, J.~Hisano, T.~Kitahara, K.~Ogawa and N.~Osamura,
[arXiv:2502.14500 [hep-ph]].

\bibitem{Ellis:1982tk}
J.~R.~Ellis, S.~Ferrara and D.~V.~Nanopoulos,
Phys. Lett. B \textbf{114} (1982), 231-234
doi:10.1016/0370-2693(82)90484-1

\bibitem{Dine:1993qm}
M.~Dine, R.~G.~Leigh and A.~Kagan,
Phys. Rev. D \textbf{48} (1993), 2214-2223
doi:10.1103/PhysRevD.48.2214
[arXiv:hep-ph/9303296 [hep-ph]].

\bibitem{Dine:1994vc}
M.~Dine, A.~E.~Nelson and Y.~Shirman,
Phys. Rev. D \textbf{51} (1995), 1362-1370
doi:10.1103/PhysRevD.51.1362
[arXiv:hep-ph/9408384 [hep-ph]].

\bibitem{Randall:1998uk}
L.~Randall and R.~Sundrum,
Nucl. Phys. B \textbf{557} (1999), 79-118
doi:10.1016/S0550-3213(99)00359-4
[arXiv:hep-th/9810155 [hep-th]].

\bibitem{Giudice:1998xp}
G.~F.~Giudice, M.~A.~Luty, H.~Murayama and R.~Rattazzi,
JHEP \textbf{12} (1998), 027
doi:10.1088/1126-6708/1998/12/027
[arXiv:hep-ph/9810442 [hep-ph]].

\bibitem{Evans:2020vil}
J.~Evans, C.~Han, T.~T.~Yanagida and N.~Yokozaki,
Phys. Rev. D \textbf{103} (2021) no.11, L111701
doi:10.1103/PhysRevD.103.L111701
[arXiv:2002.04204 [hep-ph]].

\cite{Bento:1991ez}
\bibitem{Bento:1991ez}
L.~Bento, G.~C.~Branco and P.~A.~Parada,
Phys. Lett. B \textbf{267} (1991), 95-99
doi:10.1016/0370-2693(91)90530-4

\bibitem{Valenti:2021rdu}
A.~Valenti and L.~Vecchi,
JHEP \textbf{07} (2021) no.203, 203
doi:10.1007/JHEP07(2021)203
[arXiv:2105.09122 [hep-ph]].

\bibitem{Huang:2020hdv}
G.~y.~Huang and S.~Zhou,
Phys. Rev. D \textbf{103} (2021) no.1, 016010
doi:10.1103/PhysRevD.103.016010
[arXiv:2009.04851 [hep-ph]].

\bibitem{ParticleDataGroup:2024cfk}
S.~Navas \textit{et al.} [Particle Data Group],
Phys. Rev. D \textbf{110} (2024) no.3, 030001
doi:10.1103/PhysRevD.110.030001

\bibitem{Barger:1992ac}
V.~D.~Barger, M.~S.~Berger and P.~Ohmann,
Phys. Rev. D \textbf{47} (1993), 1093-1113
doi:10.1103/PhysRevD.47.1093
[arXiv:hep-ph/9209232 [hep-ph]].

\bibitem{Hall:1993gn}
L.~J.~Hall, R.~Rattazzi and U.~Sarid,
Phys. Rev. D \textbf{50} (1994), 7048-7065
doi:10.1103/PhysRevD.50.7048
[arXiv:hep-ph/9306309 [hep-ph]].

\bibitem{Carena:1994bv}
M.~Carena, M.~Olechowski, S.~Pokorski and C.~E.~M.~Wagner,
Nucl. Phys. B \textbf{426} (1994), 269-300
doi:10.1016/0550-3213(94)90313-1
[arXiv:hep-ph/9402253 [hep-ph]].

\bibitem{Hisano:1992jj}
J.~Hisano, H.~Murayama and T.~Yanagida,
Nucl. Phys. B \textbf{402} (1993), 46-84
doi:10.1016/0550-3213(93)90636-4
[arXiv:hep-ph/9207279 [hep-ph]].

\bibitem{Asadi:2022vys}
P.~Asadi, S.~Homiller, Q.~Lu and M.~Reece,
Phys. Rev. D \textbf{107} (2023) no.11, 11
doi:10.1103/PhysRevD.107.115012
[arXiv:2212.03882 [hep-ph]].
\end{thebibliography}
\end{document}